\documentclass[preprint,amsmath,amssymb,superscriptaddress,showpacs]{revtex4}
\usepackage{graphicx}

\newcommand{\be}{\begin{equation}}
\newcommand{\ee}{\end{equation}}
\newcommand{\bea}{\begin{eqnarray}}
\newcommand{\eea}{\end{eqnarray}}

\begin{document}
\renewcommand{\thesection}{\arabic{section}}
\renewcommand{\thesubsection}{\arabic{section}.\arabic{subsection}}
\renewcommand{\thefigure}{\arabic{figure}}
\baselineskip=0.7cm

\title{Effect of Holstein phonons on the optical conductivity of gapped graphene }
\author{Kh. Jahanbani}
\affiliation{Institute for Advanced Studies in Basic Sciences
(IASBS), Zanjan, 45195-1159, Iran}
\affiliation{School of Physics, Institute for Fundamental Sciences, (IPM) 19395-5531 Tehran, Iran}
\author{Reza Asgari~\footnote{Corresponding author: Tel: +98 21 22280692; fax: +98 21 22280415.\\ E-mail address: asgari@theory.ipm.ac.ir, }}
\affiliation{School of Physics, Institute for Fundamental Sciences, (IPM) 19395-5531 Tehran, Iran}

\begin{abstract}
We study the optical conductivity of a doped graphene when a
sublattice symmetry breaking is occurred in the presence of the
electron-phonon interaction. Our study is based on the Kubo
formula that is established upon the retarded self-energy. We
report new features of both the real and imaginary parts of the
quasiparticle self-energy in the presence of a gap opening. We
find an analytical expression for the renormalized Fermi velocity
of massive Dirac Fermions over broad ranges of electron densities,
gap values and the electron-phonon coupling constants. Finally we
conclude that the inclusion of the renormalized Fermi energy and
the band gap effects are indeed crucial to get reasonable feature
for the optical conductivity.
\end{abstract}

\maketitle

\section{Introduction}
There is a considerable interest in understanding the effects on
properties of particle due to the interactions with environment,
for instance the coupling of electrons to lattice vibrations or
electron-phonon coupling. The electron-phonon coupling plays an
essential role in the theory of high temperature superconductivity
and they exist in other material such as nanotubes, C$_{60}$
molecules and other fullerenes~\cite{rmp}. Also it is important to
consider the electron-phonon coupling in transport properties.

Graphene, a single layer of carbon atoms,~\cite{Geim} is
disputable the first true two-dimensional lattices. Graphene is
thermodynamically stable and there is indeed ripple structures on
graphene sheets. Lattice displacements due to the ripple
structures are symmetric with respect to their close carbon atoms
and couple to the carrier densities. The electrons moving through
the sheet are coupled to the out-of-plane phonons and therefore
the electron-phonon coupling plays an important role in the
transport properties~\cite{akturk,basko,park}. The coupling of
electrons to out-of-plane optical phonons can be modeled by a
Holstein type coupling \cite{Holstein}. In this model the coupling
of electrons to dispersionless optical phonons is essentially
local. The electron-phonon coupling has been carefully examined
and has been shown to give rise to Kohn anomalies in the phonon
dispersion at edge points in the Brillouin zone where the phonons
can be studied by Raman
spectroscopy~\cite{piscanec1,piscanec2,pisana}. An alternative
strategy for the electron-phonon coupling measurement is based on
the analysis of the $G$-peak linewidths and its broadening.

The optical conductivity is one of the most useful tools to
investigate the basic properties of materials. Both the excitation
spectrum of materials such gaps, phonons and interband transitions
and the scattering mechanisms leave their distinct traces in
transport. It was shown that the infrared conductivity of graphene
is basically independent of the
frequency~\cite{peres2,peresijmp,peresprb78,gusynin} and
experimentally confirmed this manner~\cite{li,nair}. The effect of
electron-phonon interaction in gapless graphene has been discussed
by several authors~\cite{Stauber,calandra,tse,peres,stauperes}
directed towards understanding this effect on the optical
conductivity.

The energy spectrum of the Dirac electrons in a graphene layer
that epitaxially grown on a SiC substrate has been measured by
Zhou {\it et al.}~\cite{Zhou} and they observed an energy gap of
about 200 meV opened up in the electronic spectrum. They
attributed the opening up of the gap is due to the breaking of the
$A$ and $B$ sublattices symmetry~\cite{novoselov}. The optical
response of a gapped graphene is of important for an understanding
of optoelectronic devices. Moreover, the optical spectroscopy can
be used for measurements of the magnitude of the energy gap.

In this paper we consider the sublattice symmetry breaking
mechanism for a gap opening in a pristine doped graphene sheet and
study the impact of the electron-phonon coupling on the electronic
conductivity of the electron-doped gapped graphene using Kubo
formula at zero-temperature. We show that the renormalized
velocity is suppressed due to the electron-phonon interaction.
There is a shift in the chemical potential and we show that the
interacting chemical potential is less than the noninteracting one
due to the electron-phonon coupling. The optical conductivity is
affected by Pauli blocking below twice value of the renormalized
interacting chemical potential and gap values.

\section{Model Hamiltonian and theory}
We consider the simplest form of Hamiltonian that describes the
interaction of electron with an optical phonon mode, called the
Holstein model. The honeycomb lattice can be consider in terms of
two triangular sublattices $A$ and $B$. We consider electrons in
$\pi$-orbital of carbon atoms by using the tight-binding
Hamiltonian in addition to the effect of the electron-phonon
coupling due to localized Holstein phonons and a gap opening
procedure due to sublattice symmetry breaking~\cite{alireza}. The
total Hamiltonian in momentum space can be expressed as
    \bea
    H&=&-t\sum_{k,\sigma}[\phi(k)a_{{\bf k},\sigma}^\dagger b_{{\bf
    k},\sigma}+h.c]\nonumber\\
    &+&D\sum_{p,k,\sigma}\chi_0[a^\dagger_{{\bf p},\sigma}a_{{\bf p+k},\sigma}+
    b^\dagger_{{\bf p},\sigma}b_{{\bf p+k},\sigma}](c_{\bf k}+c_{\bf -k}^\dagger)\nonumber\\
    &+&\sum_k\omega_0c_{\bf k}^\dagger c_{\bf k}+\Delta\sum_k[a_{{\bf k},\sigma}^\dagger a_{{\bf k},_\sigma}-b_{{\bf k},\sigma}^\dagger b_{{\bf k},\sigma}]
    \nonumber\\&-&\mu_0\sum_k[a_{{\bf k},\sigma}^\dagger a_{{\bf k},\sigma}+b_{{\bf k},\sigma}^\dagger b_{{\bf k},\sigma}]
    \eea
where $a_{{\bf k},\sigma}$ or $b_{{\bf k},\sigma}$ is the fermion
annihilation operator in $k-$space on sublattice $A$ or $B$,
respectively and $t$ is the nearest neighbor hopping
parameter~\cite{castro}. The band gap, $2\Delta$ has a nonzero
value as a result of breaks the symmetry between sublattices, A
and B. We consider that the noninteracting chemical potential,
$\mu_0$ be larger than the gap value representing the
electron-doped system. The electron-phonon coupling is determined
by $D$ and furthermore $\omega_n$ denotes the fermionic Matsubara
frequency. Moreover, $c_{\bf k}$ is the annihilation phonon
operator. $\omega_0$ is the frequency of the out of plane
vibrations of the optical phonon and $\chi_0=\sqrt{\frac{\hbar}{2
M N\omega_0}}$ with M is ion's mass and N denotes the number of
unit cells. $\phi(k)=\sum_{\delta} \exp^{-i{\bf \delta}\cdot{\bf
k}}$ with $\delta$ being the vectors connecting the three nearest
neighbors on the honeycomb lattice~\cite{Mahan}. $\phi(k)$ reduces
to $\hbar v_{\rm F} k/t$ in the Dirac cone
approximation~\cite{castro}.

The matrix element of noninteracting Green's function with the gap
of the electronic spectrum is determined by following expression
   \bea
    G_{\alpha\beta}^0(k,i\omega_n)=\frac{1}{2}
    \sum_{\lambda=\pm1}\left(\delta_{\alpha,\beta}+\frac{\lambda \Upsilon_{\alpha,\beta}}{\xi_{_\Delta}(k)}\right)
    \frac{1}{i\hbar\omega_n+\mu_0-\lambda\xi_{_\Delta(k)}},\nonumber
    \label{GAA}
    \eea
in which $\alpha, \beta=A, B$ and we have defined parameters
$\Upsilon_{AA}=-\Upsilon_{BB}=\Delta$,
$\Upsilon_{AB}=\Upsilon^{*}_{BA}=-t\phi(k)$. The quasiparticle
excitation energy is
$\xi_{_\Delta}(k)=\sqrt{t^2|\phi(k)|^2+\Delta^2}$. Note that at
zero-temperature $\mu_0=\xi_{_\Delta}(k_{\rm F})$ with $k_{\rm F}$
is the Fermi momentum of charge carriers.

An exact evaluation of the self-energy is only possible in some
special cases. The matrix elements of the self-energy calculated
to the lowest order in the electron-phonon interaction and is
defined as \be \Sigma_{\alpha,\beta}(i\omega_n,p)=-k_B
T\sum_{k,\nu}
    D^2\chi_0^2D^{(0)}(k,i\nu)
    G^{(0)}_{\alpha,\beta}(p-q,i\omega_n-i\nu)
\ee where $G^{(0)}_{\alpha,\beta}$ and $D^{(0)}$ are the
zero-order electron and phonon Green's functions,
respectively~\cite{Mahan,Jonson}. In Holstein phonons, $\chi_0$
and $D^{(0)}(k,i\nu)$ are momentum independent and thus the phonon
propagator is simplified by \be
D^{(0)}(k,i\nu)=-2\omega_0/(\nu^2+\omega_0^2). \ee
 We restrict our calculations to
the lowest order self-energy that is sufficient if Migdal's
theorem, states that vertex corrections in the electron-phonon
interaction can be neglected if the typical phonon frequencies are
sufficiently smaller than the electronic energy scale, is valid.
Therefore, we can neglect the vertex corrections since the
self-energy is $k$-independent. Using the contour integration, we
can perform the summation over the bosonic frequency in the
expression of the self-energy and finally the self-energy yields
as
    \bea
    \Sigma_{AA}(i\omega_n)&=&\frac{D^2\chi_0^2}{2}\sum_{k,\lambda=\pm1}
    \left(1+\frac{\lambda\Delta}{\xi_{_\Delta(k)}}\right)\times\nonumber\\
    &&\left\{\frac{N_0+n_F(\lambda\xi_{_\Delta(k)}-\mu_0)}
    {i\hbar\omega_n+\hbar\omega_0-\lambda\xi_{_\Delta(k)}+\mu_0}
    \!+\!\frac{N_0+1-n_F(\lambda\xi_{_\Delta(k)}-\mu_0)}
    {i\hbar\omega_n-\hbar\omega_0-\lambda\xi_{_\Delta(k)}+\mu_0}\right\},
    \label{general-self-energy}
    \eea
where $N_0=1/(e^{\hbar\omega_0/k_B T}-1)$ and $n_F(x)$ denotes the
Fermi-Dirac distribution function. To calculate
$\Sigma_{BB}(i\omega_n)$, the gap value $\Delta$ might be replaced
by $-\Delta$ in Eq.~\ref{general-self-energy}. It should be noted
that $\Sigma_{AB}(i\omega_n)=\Sigma_{BA}(i\omega_n)=0$ in the
Dirac cone approximation. The explicit expression of the
self-energy will be computed in the following.

\subsection{Finite doping with a gap opening}

We consider the low excited electron energy where the
noninteracting electron spectrum energy is given by $\sqrt{(\hbar
v_{\rm F} k)^2+\Delta^2}$~\cite{alireza}. To evaluate the
zero-temperature retarded self-energy evaluated at the Fermi
surface, we integrate Eq.~\ref{general-self-energy} over $k$ and
then decompose the results into
$\Sigma^j(\omega)=\Sigma^j_0(\omega)+\Delta \Sigma^{j}(\omega)$
where
    \bea\label{reself1}
    \Re e\Sigma^j_0(\omega)=\frac{\hbar A_c}{2\pi}
    (g\omega_0)^2
    \{-\frac{\omega_j}{v_F^2}
    \ln|\frac{\Delta^2+\hbar^2v_F^2k_c^2}
    {(\hbar\omega+\mu_0)^2-(\hbar\omega_0+\Delta)^2}|
    +\frac{\omega_0}{v_F^2}
    \ln|\frac{\omega_{+}+\omega_0}{\omega_{-}-\omega_0}|\}
    \eea
    \bea\label{imself1}
    \Im m\Sigma^j_0(\omega)&=&-\pi\frac{\hbar A_c}{2\pi}
    (g\omega_0)^2
    \{-\frac{\omega_j+\omega_0}{v_F^2}
    \Theta(-\omega_{+}-\omega_0)
    \Theta(\hbar\omega+\mu_0+\hbar\omega_0+\sqrt{\hbar^2v_F^2k_c^2+\Delta^2})\nonumber\\
    &+&\frac{\omega_j-\omega_0}{v_F^2}
    \Theta(\omega_{-}-\omega_0)
    \Theta(-\hbar\omega-\mu_0+\hbar\omega_0+\sqrt{\hbar^2v_F^2k_c^2+\Delta^2})\}\nonumber
    \eea
here $\omega_{j}=\omega+(\mu_0+j\Delta)/\hbar$ with $j=+1(-1)$
refers to sublattice $A$($B$). $k_c$ is the ultraviolet cut-off
momentum~\cite{Stauber} and finally the coupling constant
$g=\sqrt{N}D \chi_0/\omega_0$ being the order of unity. The area
of the unit cell is $A_c=a^2 3\sqrt{3}/2$ with $a=1.42${\AA}. The
extra terms take the following form as
    \bea\label{reself2}
    \Re e\Delta\Sigma^j(\omega)=\frac{\hbar A_c}{2\pi}
    (g\omega_0)^2
    \{-\frac{\omega_j}{v_F^2}\ln|\frac{(\omega+\omega_0)(\omega_{-}-\omega_0)}
    {(\omega-\omega_0)(\omega_{-}+\omega_0)}|
    -\frac{\omega_0}{v_F^2}
    \ln|\frac{\omega^2-\omega_0^2}{(\omega_{-})^2-\omega_0^2}|\}
    \eea
    \bea\label{imself2}
    \Im m\Delta\Sigma^j(\omega)=-\pi\frac{\hbar A_c}{2\pi}
    (g\omega_0)^2
    \{&&\!\!\!\!\!\!\frac{\omega_j+\omega_0}{v_F^2}~\Theta(\omega_{-}+\omega_0)\Theta(-\omega-\omega_0)\nonumber\\
    &-&\frac{\omega_j-\omega_0}{v_F^2}
    \Theta(-\omega+\omega_0)~\Theta(\omega_{-}-\omega_0)\}
    \eea
If $\Delta=0$, the self-energy reduces to massless Dirac graphene
which addressed in Ref~\cite{Stauber}. Therefore, we have
generalized the retarded self-energy expression to gapped
graphene. Once the retarded self-energy is obtained, the
quasiparticle properties of system due to the interaction of the
electron-phonon can be calculated. The renormalized electronic
spectrum is given by the Dyson equation as $E_{\bf
k}=\xi_{\Delta}(k)+\Re e\Sigma(E_{\bf k})$. Notice that according
to the Dyson equation, we might distinguish the noninteracting
chemical potential from the chemical potential of the interacting
system due to the fact that $\Re e \Sigma(\omega)$ is not vanished
for doped graphene when $\omega$ tends to zero. We thus have
    \be
    \mu=\mu_0+\Re e\Sigma(\omega)|_{\omega=0}~.
    \ee

The renormalized velocity, on the other hand, is given by
    \be
    \frac{v^\star}{v_{\rm F}}=\frac{\hbar v_{\rm F} k/\xi_{\Delta}(k)+(v_{\rm F}\hbar)^{-1}\partial_{k}
    \Re e\Sigma(k, \omega)}{1-\hbar^{-1}\partial_{\omega}
    \Re e\Sigma(k, \omega)}|_{\omega=0, k=k_F}
    \ee
within the Dyson scheme~\cite{Mahan}. The self-energy is
independent of the momentum, accordingly its $k$-derivative is
zero. Consequently, the renormalized velocity is obtained
analytically
    \bea
    \frac{v_{\rm F}}{v^*(1+(\Delta/\hbar v_{\rm F} k_{\rm F})^{1/2})}
    &=&1+\left(\frac{g{\omega}_0}{v_{\rm F}{k}_c}\right)^2\{
    \ln|\frac{({\Delta}^2+({\hbar v_{\rm F}k_c})^2)}
    {(({\mu_0}+{\hbar\omega}_0)^2-{\Delta}^2)}|\nonumber\\
    &-&({\mu_0}+j{\Delta}+\hbar{\omega}_0)
    \frac{2({\mu_0}+\hbar{\omega}_0)}
    {({\mu_0}+\hbar{\omega}_0)^2-{\Delta}^2}
    +2\frac{{\mu_0}+j{\Delta}}{\hbar{\omega}_0}\}.
\eea

\subsection{Optical Conductivity}

The optical conductivity can be calculated from the Kubo
formalism. To this end,
 we need to obtain the current
operator which is a composition of the paramagnetic and
diamagnetic terms, i.e.
$j_\alpha=j^P_\alpha+j^D_{\alpha\beta}{\emph{A}}_\beta $. We do
need to modify the hopping parameter in the presence of an
electromagnetic field~\cite{Stauber} and then expand it up to the
second order in the vector potential
$\overrightarrow{\emph{A}}(t)$. The current operator expressions
do not change in the presence of the gap value and therefore by
assuming that the electric field is in the direction of $x$-axis,
we have
   \be
   j_x^P=-i\xi\sum_{\sigma,k}
   [(\phi(k)-3)a_{\sigma}^\dagger(k) b_{\sigma}(k)
    -(\phi^\star(k)-3)a_{\sigma}(k) b_{\sigma}^\dagger(k)]
   \ee
where $\xi=t e a/\hbar$ and then the Kubo formula for conductivity
is given by
    \be
    \sigma_{xx}(\omega)=\frac{<j_x^D>}{i A_s (\omega+i\eta)}+
    \frac{\Lambda_{xx}(\omega+i\eta)}{i\hbar A_s (\omega+i\eta)}
     \ee
where $A_s$ is the area of sample and
$\Lambda_{xx}(i\omega_n)=\int_0^{\hbar/ k_B T }d\tau
    e^{i\omega_n\tau}<T_\tau j_x^P(\tau)j_x^P(0)>
$~\cite{Mahan}. We have ignored vertex corrections in the Kubo
formula since we worked in nearly highly electron doped graphene
for which the Dirac cone approximation is applicable. It was shown
that the vertex corrections is essential for the low density
carriers of the DC conductivity of graphene.~\cite{cappelluti}
After a lengthy but straightforward algebra, we find
\bea\label{con}
    \Im m\Lambda_{xx}(\omega)\!&=&\!\!\xi^2
    \frac{A_s}{8\pi}\!\int_0^{k_c}\! kdk\int_{-\infty}^{\infty}\frac{d\epsilon}{2\pi}
    (n_F(\epsilon+\omega)-n_F(\epsilon))\nonumber\\
    &\times&\left\{(2t^2|\phi(k)|^4)
    A_{AB}(k,\epsilon)A_{AB}(k,\epsilon+\omega)\right.\nonumber\\
    &+&(9-|\phi(k)|^2)~[A_{AA}(k,\epsilon)A_{BB}(k,\epsilon+\omega)\nonumber\\
    &+&A_{BB}(k,\epsilon)A_{AA}(k,\epsilon+\omega)]\}
    \eea

where the spectral functions are the imaginary part of Green's
function which take the following forms:
    \bea
    A_{\alpha,\beta}=-2\Im m\{\frac{\Phi_{\alpha,\beta}}{{(\Omega_{+}-\Sigma_{BB}(i\omega_n))
    (\Omega_{-}-\Sigma_{AA}(i\omega_n))-t^2|\phi(k)|^2}}\}.\nonumber
    \eea

Here
 \bea
 \Omega_{\pm}=i\hbar\omega_n+\mu \pm \Delta,~~~~~~~~~
 \Phi_{AA(BB)}=\Omega_{+(-)}-\Sigma_{BB(AA)}(i\omega_n)\nonumber
 \eea
and $\Phi_{AB}=\Phi_{BA}=1$. The integral over $k$ in
Eq.~\ref{con} can be performed analytically and accordingly one
dimensional integral will be needed to be calculated numerically.
Note that the interacting chemical potential is used instead of
the noninteracting one because of the nonzero value of
$\Sigma_j(0)$ . It should be noted that by setting $\Delta=0$, the
optical conductivity results are different with the results given
in Ref.~\cite{Stauber} due to the fact that we have implemented
the interacting Fermi energy in the formalism.

\section{Numerical Results}
We have considered the system with the phonon energy being
$\hbar\omega_0=0.2~eV$~\cite{peres}. Although the order of
coupling constant is unity,
 we consider a
larger value to seek its effect better. We have found that the
value of the quasiparticle properties for sublattices $A$ and $B$
are different at most about $0.8\%$ due to the gap opening. We
will then present only the results of the sublattice $A$.

In Fig.~1, we have shown the results of the real and imaginary
parts of the retarded self-energy for the electron-doped system,
($\mu>\Delta$) at $n=5 \times 10^{12}$cm$^{-2}$. $\Im m
\Sigma(\omega)$ vanishes in $|\omega|<\omega_0$ at which point it
jumps up to a finite value because only then can a quasiparticle
decay by boson emission. It drops towards the zero for
$\omega<-\omega_0$ and then increases linearly showing a marginal
type physics which happens in the Coulomb electron interactions in
undoped graphene~\cite{polini}. Notice that $\Im m \Sigma$ is not
symmetric with respect to change of the sign of frequency. In
addition, $\Im m \Sigma(\omega)$ vanishes when
$|\hbar\omega+\hbar\omega_0+\mu_0|<\Delta$ due to the effect of
the gap opening and $\Im m\Sigma$ tends to zero at
$-\hbar\omega_0-\mu_0$ for $\Delta=0$. These behaviors can be
determined explicitly from expressions given by Eqs.~\ref{imself1}
and \ref{imself2}.

In Fig.~1b we can see logarithmic type singularities~\cite{dogan}
at $\omega=\pm \omega_0$ and
$\omega=-\omega_0-(\mu_0-\Delta)/\hbar$ for the results of $\Re e
\Sigma$. The extra singular behavior is due to the gap effect. It
should be noted that the singularity at $\omega=\pm \omega_0$
would be washed out if a momentum dependence of phonon spectra is
used. In addition, there is a cancelation of the logarithmic
singularity at $\omega=-\omega_0-(\mu_0+\Delta)/\hbar$. The
logarithmic singularity can be determined to the argument of the
logarithm in Eqs.~\ref{reself1} and \ref{reself2}. We have
obtained an expression for the interacting density of states too
through the spectral function. The singularities manner lead to
kink structures in the interacting electronic density of states.
In the results, there are three kink structures in the interacting
density of states where one of them is associated to the gap. The
kink structures would affect to physical quantities and transport
properties through the interacting electronic density of states.

The renormalized velocity as function of the densities, gap values
and the coupling constants are shown in Fig.~2. The renormalized
velocity is suppressed due to the electron-phonon interaction and
the gap values too. We have found a nonmonotonic behavior of $v^*$
with respect to the electron density when the gap value increases
and results are shown in Fig.~2b. At small gap values, $v^*$
decreases with increasing density however it changes behavior at
large gap values and behaves like conventional two-dimensional
electron systems. Therefore, we expect that the electron-phonon
interaction renormalized the electronic quantities at the Fermi
surface by a factor $v^*/v_{\rm F}$.

The optical conductivity scaled by $\sigma_0=e^2/4\hbar$ as a
function of energy for different values of (a) the coupling
constants and (b) the gap values are shown in Fig.~3. First of
all, $\sigma$ tends to a minimum value at $\omega_0$. Moreover it
basically increases around $\omega>\omega_0$ due to the
contribution of the Holstein phonon sideband. In the case of
noninteracting electron-phonon system, $\sigma$ has a sharp
structure, step function manner, at $2\mu_{0}$ due to interband
transitions and the conductivity increases by a factor of two,
$\sigma=2\sigma_0$ at $\omega=2\Delta$ and finally at higher
frequencies decreases and approaches to $\sigma_0$~\cite{gusynin}.
By switching interaction on, the chemical potential becomes weaker
and consequently the position of the sharp structure changes to
$2\mu$ which is smaller than $2\mu_{0}$. This behavior is clearly
shown in the Fig.~3 which did not consider in results discussed in
Ref.~\cite{Stauber}. At $g=0$, the conductivity is larger than
$\sigma_0$ about $2\mu$ and then tends to $\sigma_0$ in gapped
graphene. However, $\sigma$ always remains smaller than $\sigma_0$
in gapless graphene. The gap dependence on the optical
conductivity is shown in Fig.~3b. First, the gap opening makes the
chemical potential bigger therefore the sharp structure in the
$\sigma$ tends to larger $\omega$ values. Second, the scattering
mechanism increases by increasing the electron densities and then
the optical conductivity changes and becomes smaller.

Another point of interest for experiments is the density
dependence ( in units of 10$^{12}$ cm$^{-2}$) of the optical
conductivity ( Fig.~4) as a function of frequency at $2\Delta=0.2$
eV. Note that the noninteracting chemical potential values
associated to the electron densities used in Fig.~4 are
$\mu_0=0.154, 0.279, 0.382$ and $0.831$ eV, respectively with
giving $\Delta=0.1$ eV. The optical conductivity increases by
increasing the electron density around $\omega_0$ however $\sigma$
decreases faster by increasing the density at high frequency. The
sharp structure of the optical conductance tends to higher
frequency by increasing the electron density. The sharp position
occurs at $2\mu$ which is always smaller than $2\mu_0$ for the
same system.

\section{Conclusion}
we have calculated the optical conductivity of gapped graphene,
including the effect of the lowest order self-energy diagram due
to the electron-phonon interaction by Holstein Hamiltonian. We
have reported an extra logarithmic singular behavior associated to
gap value in the real part of the self-energy. We have found the
density, gap value and the electron-phonon coupling dependence of
the renormalized velocity and the interacting chemical potential.
The optical conductivity is affected by these physical quantities
and Pauli blocking below twice value of the renormalized chemical
potential and the gap values. We conclude that the inclusion of
the renormalized Fermi energy and the band gap affects are indeed
crucial to get reasonable feature for the optical conductivity.
The gap dependence of the optical conductivity would be verified
by experiments.

\section{acknowledgments}
R. A thank S.G. Sharapov for stimulating discussion. We are
grateful A. Qaiumzadeh for useful comments. We thank Centro de
Ciencias de Benasque, Spain where this work was completed.

{\it Note added-} In final stage of preparing this manuscript, we
became aware of a related work for gapless graphene
\cite{carbotte}.

\newpage

\begin{figure}[h]
\centerline{\includegraphics[width=.41\columnwidth]{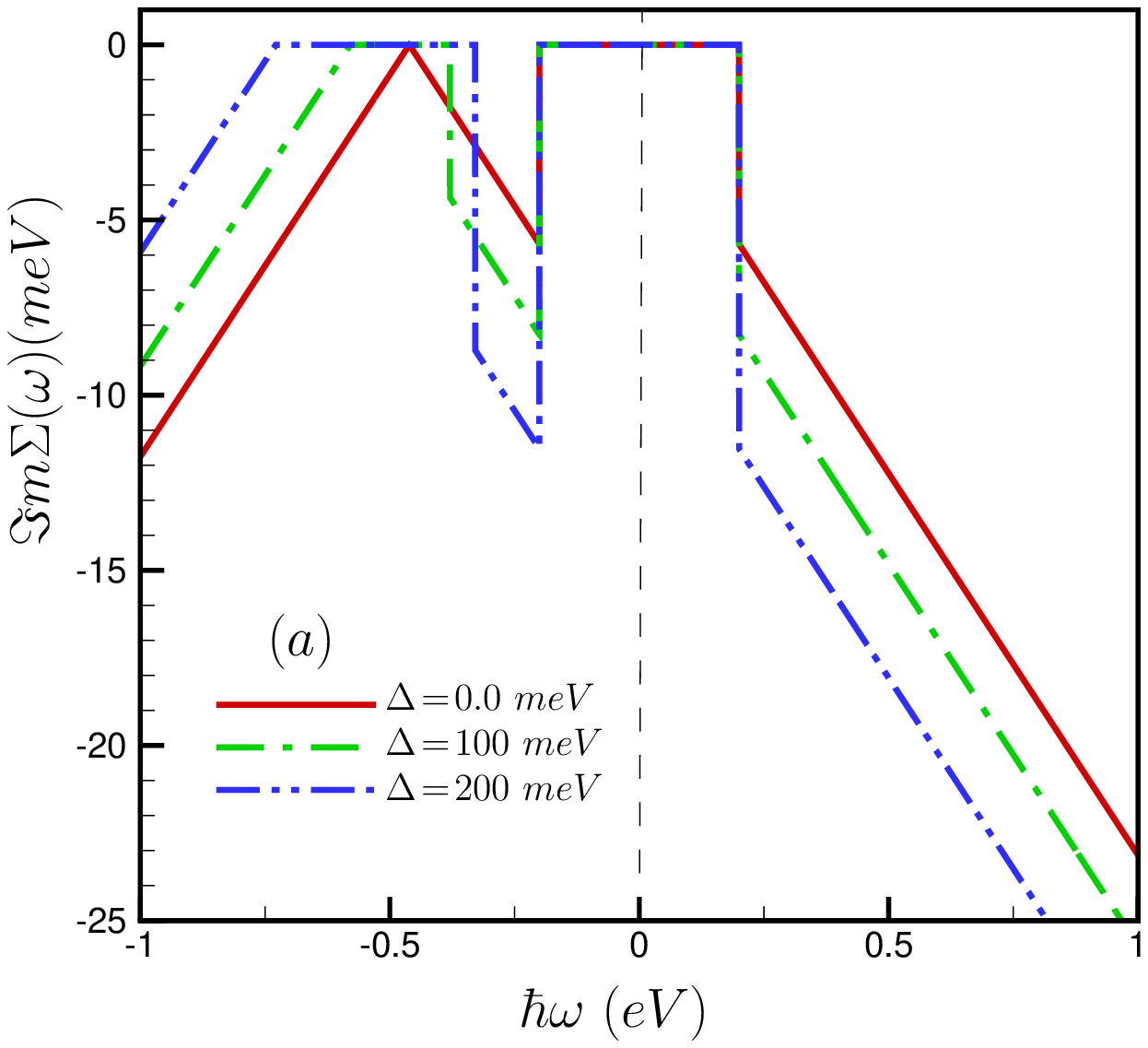}
\includegraphics[width=.41\columnwidth]{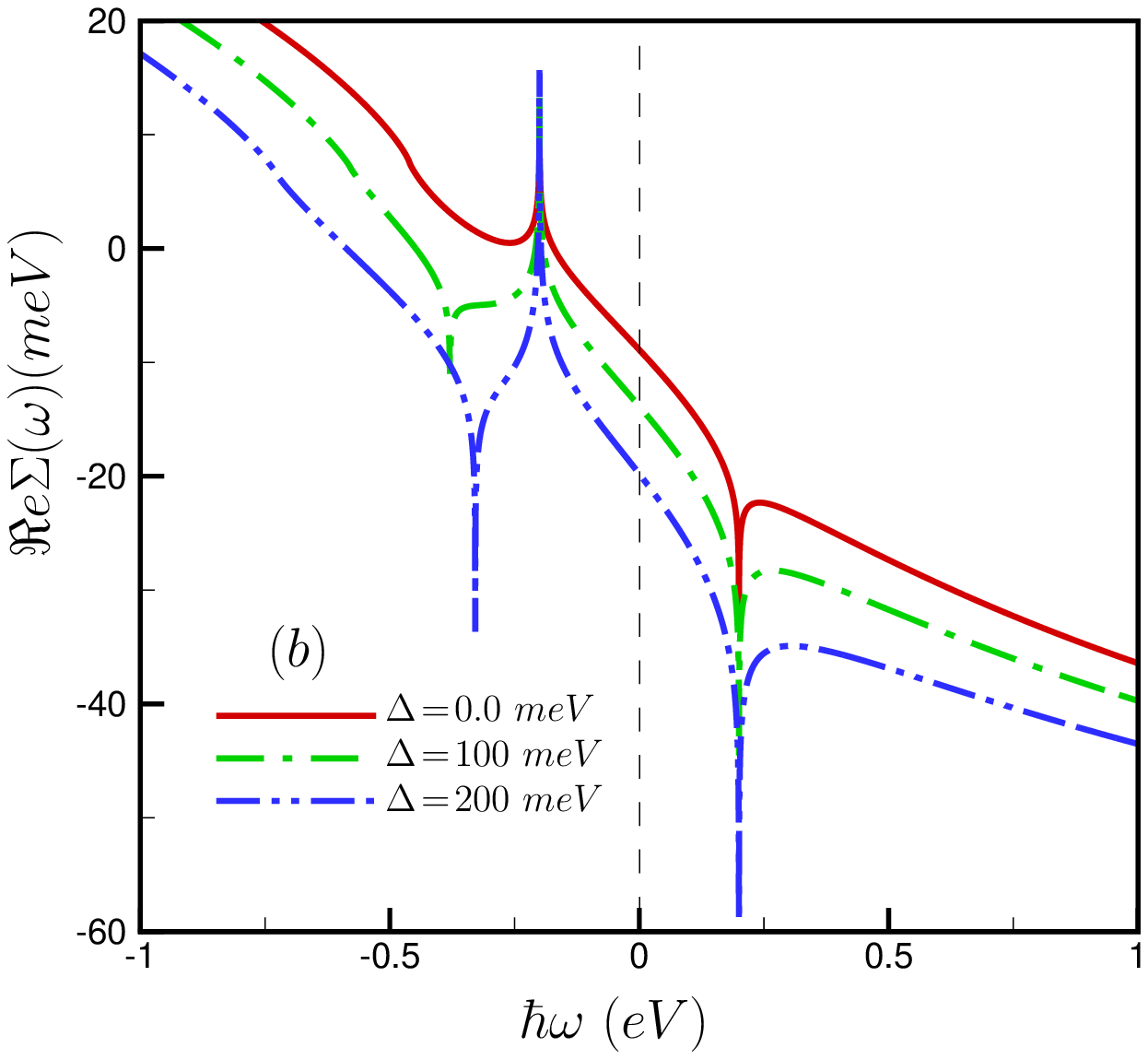}}
\caption{(Color online) Imaginary (a) and real (b) parts of the
self-energy as a function of energy evaluated at Fermi energy for
different gap values at the coupling constant $g=3.0$ and density
$n=5.0 \times 10^{12}$ cm$^{-2}$.}
\end{figure}

\begin{figure}[h]
\centerline{\includegraphics[width=.41\columnwidth]{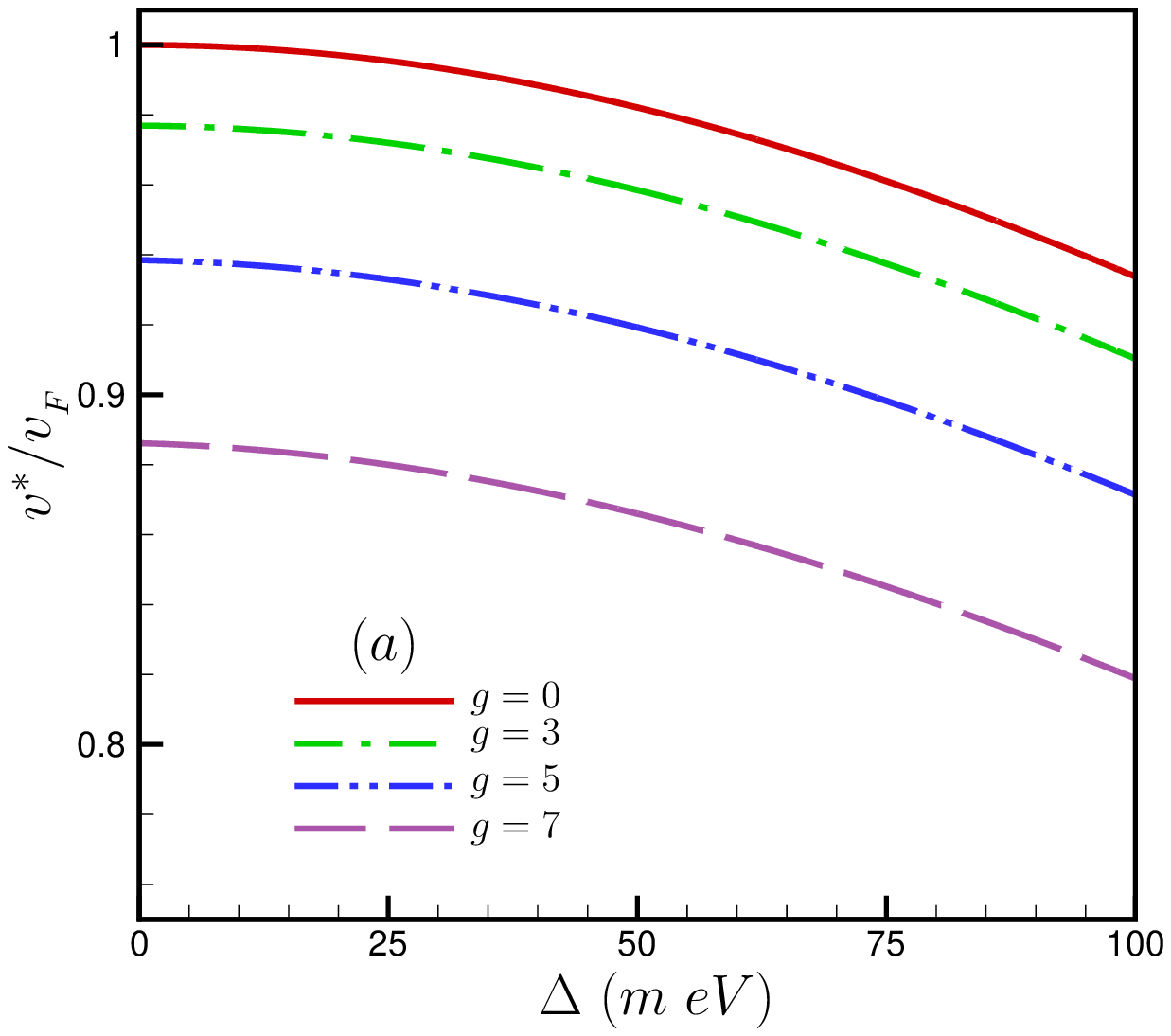}
\includegraphics[width=.41\columnwidth]{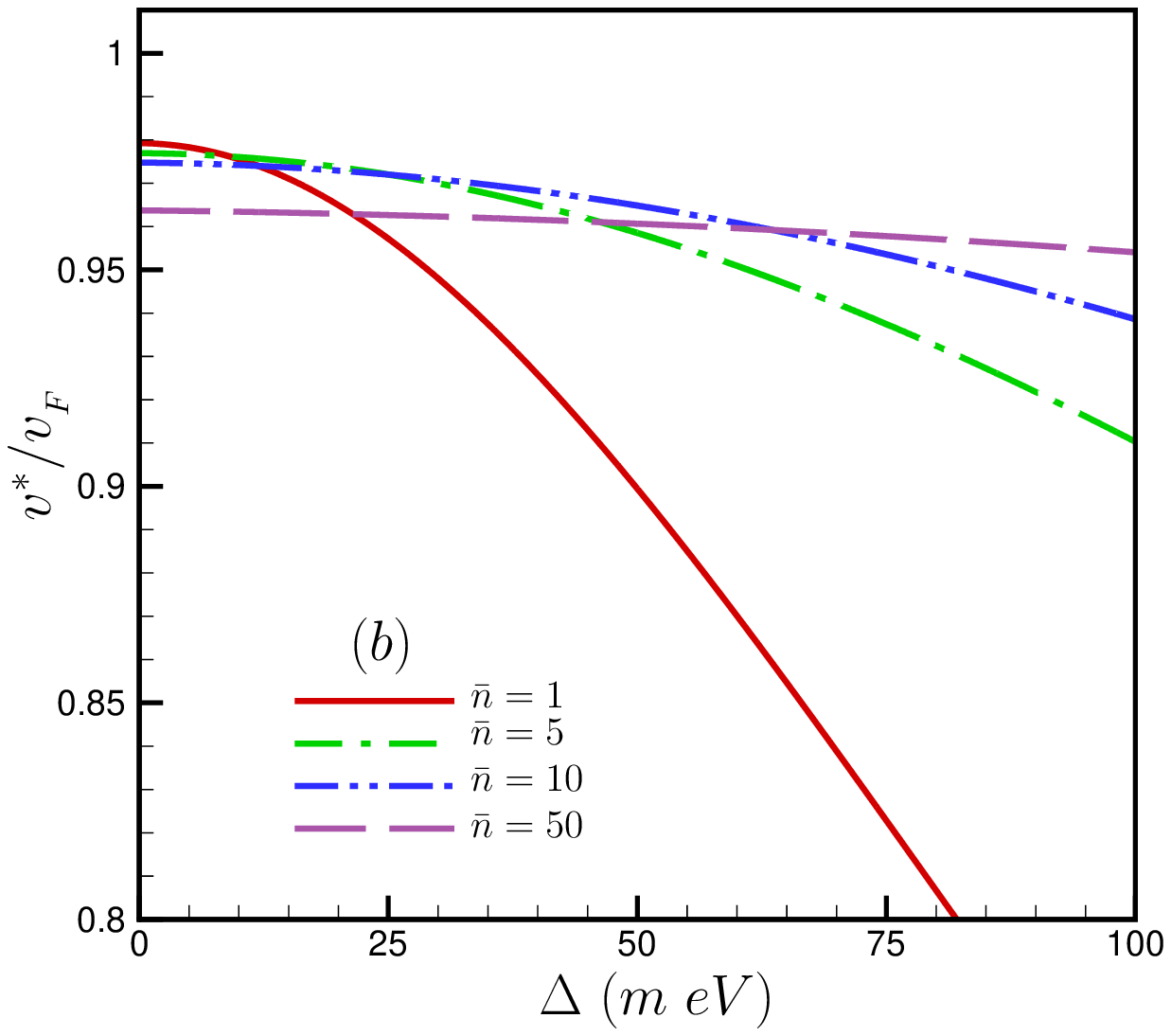}}
\caption{(Color online) Renormalized electron velocity for (a) the
different values of coupling constants at $n= 5.0\times10^{12}$
cm$^{-2}$, (b) the different value of density (in units of
$10^{12}$ cm$^{-2}$) at $g=3.0$.} \label{fig3}
\end{figure}

\begin{figure}[h]
\centerline{\includegraphics[width=.41\columnwidth]{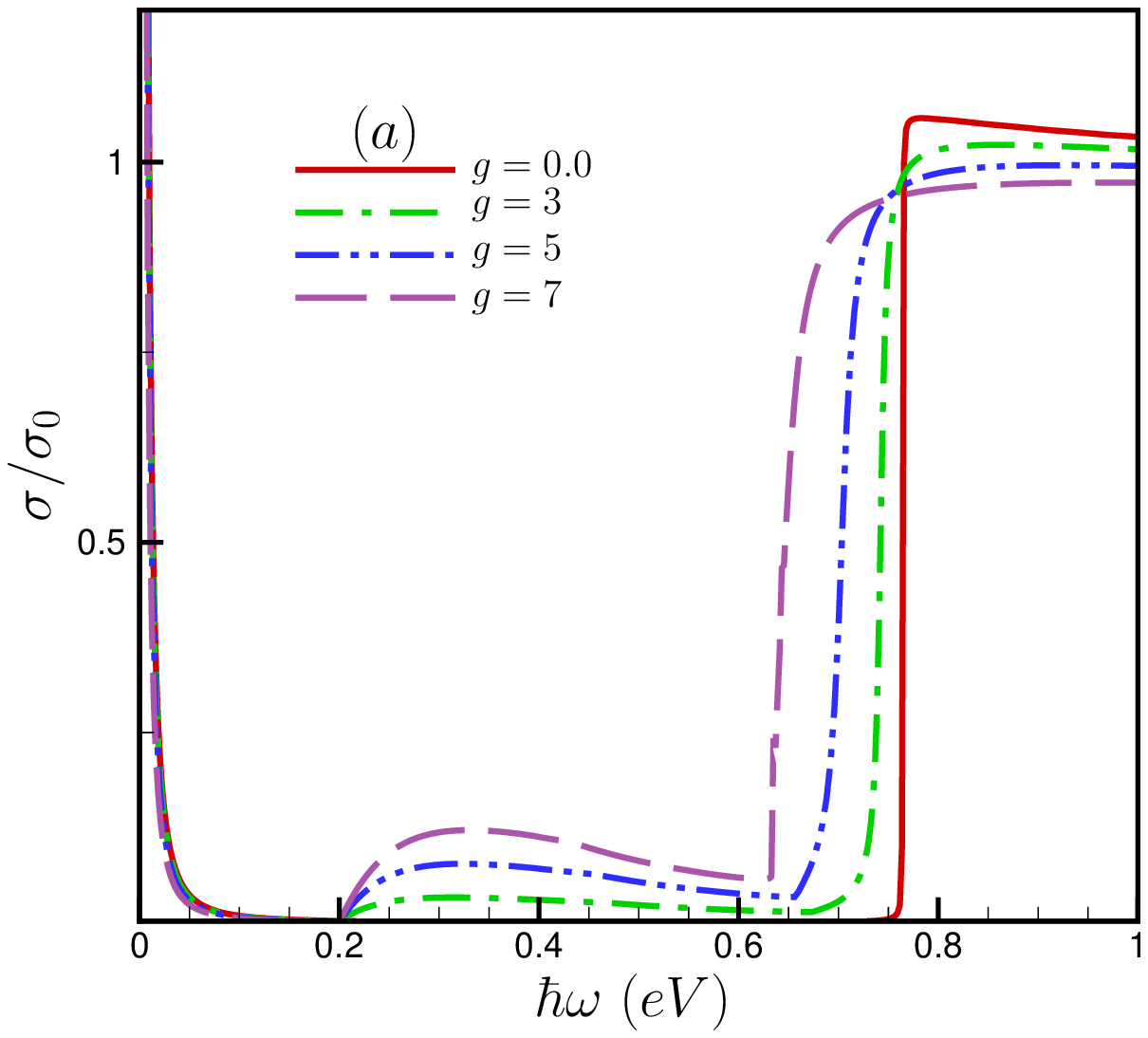}
\includegraphics[width=.41\columnwidth]{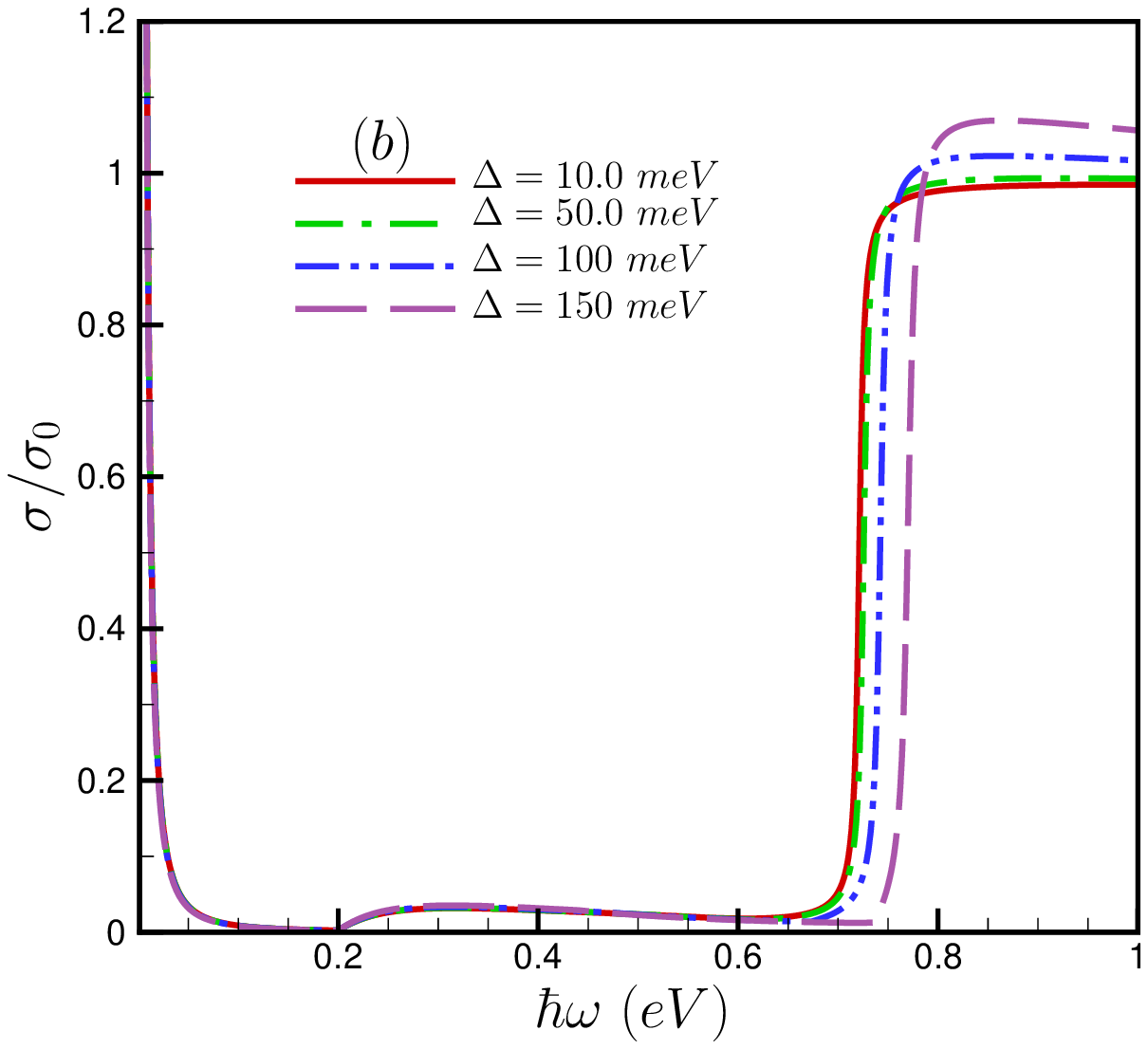}}
\caption{(Color online) Optical conductivity as a function of
energy for (a) the different values of coupling constant at
$\Delta=0.1~eV$ and (b) the different value of $\Delta$ at
$g=3.0$. We consider $n=1\times10^{13}~cm^{-2}$.}
\end{figure}

\begin{figure}[h]
\centerline{\includegraphics[width=.5\columnwidth]{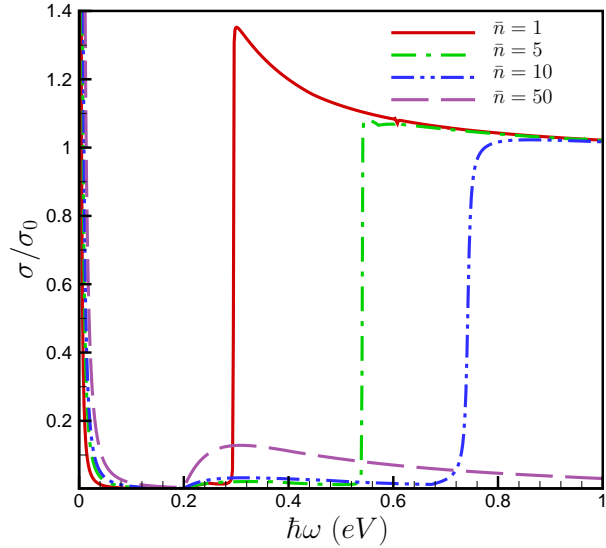}}
\caption{(Color online) Optical conductivity as a function of
energy for different values of density ( in unites of $10^{12}$
cm$^{-2}$) at $g=3.0$ and $\Delta=0.1~eV$}
\end{figure}

\end{document}